\documentclass[12pt,a4paper]{article}

\usepackage{graphics}
\usepackage{latexsym}
\usepackage{amsmath}

\title{Three generation neutrino mixing is compatible with 
all experiments}
\author{B. Hoeneisen and C. Mar\'{\i}n}
\date{\small{Universidad San Francisco de Quito} \\
	2 February 2000 \\}
\begin{document}
\maketitle

\begin{abstract}
\noindent
We consider the minimal extension of the Standard Model
with three generations of massive neutrinos that
mix. We then determine the
parameters of the model that satisfy all experimental
constraints.

\end{abstract}

Three observables in disagreement with the Standard Model of Quarks and
Leptons are: i) A deficit of electron-type solar neutrinos; ii) A deficit
of muon-type atmospheric neutrinos; and, possibly,
iii) The observation of the
apearance of $\bar{\nu}_e$  in a beam of $\bar{\nu}_\mu$ by the LSND
Collaboration. 
The invisible width of the $Z$ implies
that the number of massless, or light Dirac, or 
light Majorana neutrino species is
$N_\nu = 2.993 \pm 0.011$.\cite{rev_part_phys}
To account for these observations we consider the minimal
extension of the Standard Model with three massive
neutrinos that mix. The neutrino interaction eigenstates
$\nu_l$ are superpositions of the neutrino mass eigenstates $\nu_m$:
\begin{equation}
\vert\nu_{l}\rangle=\sum_{m} U_{lm}\vert\nu_{m}\rangle
\label{neutrino}
\end{equation}
We consider the \textquotedblleft{standard}" 
parametrization of the unitary matrix
$U_{lm}$\cite{rev_part_phys}:
\begin{equation}
\left(\begin{array}{c}
\nu_e \\
\nu_\mu \\
\nu_\tau
\end{array}
\right)
=
\left(\begin{array}{ccc}
c_{12}c_{13} & s_{12}c_{13} & s_{13}e^{-i\delta}\\
-s_{12}c_{23}-c_{12}s_{23}s_{13}e^{i\delta} &
c_{12}c_{23}-s_{12}s_{23}s_{13}e^{i\delta} &
s_{23}c_{13} \\
s_{12}s_{23}-c_{12}c_{23}s_{13}e^{i\delta} &
-c_{12}s_{23}-s_{12}c_{23}s_{13}e^{i\delta} &
c_{23}c_{13}
\end{array}
\right)
\left(\begin{array}{c}
\nu_1 \\
\nu_2 \\
\nu_3
\end{array}
\right)
\label{matrix}
\end{equation}
where $c_{ij} \equiv cos{\theta_{ij}}$, $s_{ij} \equiv sin{\theta_{ij}}$,
$0 \leq \theta_{ij} \leq \frac{\pi}{2}$ and $-\pi \leq \delta < \pi$.
The probability that an ultrarelativistic neutrino produced as
$\nu_l$ decays as $\nu_{l'}$ is\cite{rev_part_phys}:
\begin{equation}
P(\nu_l \rightarrow \nu_{l'}) = \vert \sum_m U_{lm} 
exp(-iLM_m^2 / 2E) U_{{l'}m}^* \vert ^2 =
P(\bar{\nu}_{l'} \rightarrow \bar{\nu}_l)
\label{probability}
\end{equation}
where $E$ and $L$ are the energy and traveling distance of $\nu_l$, and
$M_m$ is the mass of $\nu_m$. We choose $M_1 \le M_2 \le M_3$.
This extension of the Standard Model introduces six parameters:
$s_{12}$, $s_{23}$, $s_{13}$, $\delta$, and two mass-squared 
\mbox{differences},
\textit{e.g.} $\Delta M_{21}^2 \equiv M_2^2 - M_1^2$ and
$\Delta M_{32}^2 \equiv M_3^2 - M_2^2$. We vary these parameters
to minimize a $\chi^2$. This $\chi^2$ has $14$ terms obtained 
from the solar neutrino
data summarized in Table \ref{solar}, the atmospheric neutrino data shown 
in Table \ref{atmospheric}, and the LSND data\cite{LSND}:
$P(\bar{\nu}_\mu \rightarrow \bar{\nu}_e) =
\frac{1}{2} \sin^2(2 \theta) =
0.0031 \pm 0.0013$ for
$L$[km]$/E$[GeV]\mbox{$=[P(\bar{\nu}_\mu \rightarrow \bar{\nu}_e)]^{1/2}/
1.27 \cdot \Delta M^2$[eV$^2$]}$\approx 0.73$
(here $\sin^2(2 \theta)$ corresponds to 
\textquotedblleft{large}" $\Delta M^2$,
and $\Delta M^2$ corresponds to $\sin^2(2 \theta)=1$, see discussion in
\cite{rev_part_phys}). Because  one author\cite{hill} of the LSND Collaboration
is in disagreement with the conclusion, and because the result has
not been confirmed by an independent experiment, we multiply the error by
$1.5$ and take 
$P(\bar{\nu}_\mu \rightarrow \bar{\nu}_e) = 0.0031 \pm 0.0020$.
We require that the astrophysical, reactor and accelerator
limits be satisfied. The most stringent of these limits are listed in
Table \ref{reactor}.

\begin{table}
\begin{center}
\begin{tabular}{|l|l|c|c|c|}
\hline
Experiment & Energy & Observed   & SSM predic-  & Ratio \\
           & [MeV] & flux (SNU) & tion (SNU)   &       \\
\hline
Homestake\cite{homestake} & $0.87$  & $2.56 \pm 0.23$ & $7.7^{+1.2}_{-1.0}$ & $0.33 \pm 0.05$ \\
Sage\cite{sage} & $0.233-0.4$ & $67 \pm 8$     & $129^{+8}_{-6}$   & $0.52 \pm 0.07$ \\
Gallex\cite{gallex} & $0.233-0.4$ & $78 \pm 8$  & $129^{+8}_{-6}$   & $0.60 \pm 0.07$ \\
Kamiokande\cite{K} & $7-13$ & $2.80 \pm 0.38$ & $5.2^{+1.0}_{-0.7}$ &$0.53
\pm 0.11$ \\ 
Super-Kam.\cite{superK_solar} & $6-13$ & $2.42^{+0.12}_{-0.09}$ &
$5.2^{+1.0}_{-0.7}$ & $0.46 \pm 0.08$ \\
\hline
\end{tabular}
\end{center}
\caption{Observed solar electron-type neutrino flux, 
compared to the Standard Solar Model (SSM) predictions 
asuming no neutrino mixing\cite{bahcall}, and their ratio.
The Solar Neutrino Unit (SNU) is $10^{-36}$ captures per 
atom per second.
For Kamiokande (Super Kamiokande) the flux is in units of
$10^6$cm$^{-2}$s$^{-1}$ at Earth above $7$MeV ($6.5$MeV).}
\label{solar}
\end{table}

\begin{table}
\begin{center}
\begin{tabular}{|l|c|c|}
\hline
$L/E$ [km/GeV] & $R_e$ & $R_\mu$ \\ 
\hline
10       & $1.20 \pm 0.15$ & $1.00 \pm 0.15$  \\
100      & $1.20 \pm 0.15$ & $0.85 \pm 0.12$  \\
1000     & $1.20 \pm 0.15$ & $0.70 \pm 0.10$  \\
10000    & $1.20 \pm 0.15$ & $0.60 \pm 0.08$  \\
\hline
\end{tabular}
\end{center}
\caption{Ratio of the numbers of observed and predicted 
electron-type and muon-type neutrinos as a
function of the flight length-to-energy ratio
as measured by the Super-Kamiokande Collaboration.\cite{superK_atm}
Because of the uncertainty on the absolute neutrino
flux and because $R_e$ is observed to be independent of $L/E$, we divide
the numbers in this table by $1.15$ so that $R_e \approx 1$.}
\label{atmospheric}
\end{table}

\begin{figure}
\begin{center}
\vspace*{-4.5cm}
\scalebox{0.4}
{\includegraphics[0in,1in][8in,9.5in]{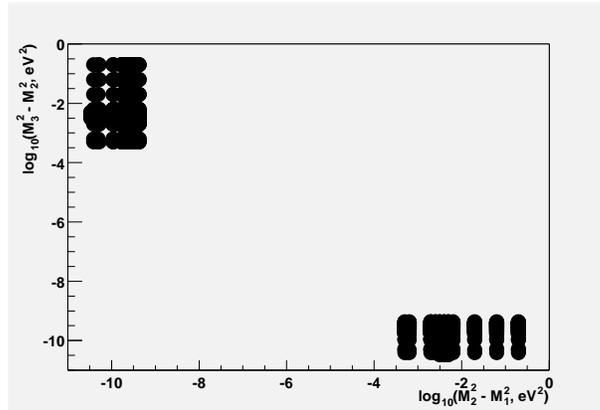}}
\vspace*{0.7cm}
\caption{The mass-squared differences
$(M_2^2-M_1^2,M_3^2-M_2^2)$ lie within
the dots with $90\%$ confidence.}
\label{islands}
\end{center}
\end{figure}

The $\chi^2$ has $8$ degrees of freedom ($14$ terms minus $6$ parameters).
Varying the parameters we obtain minimums of $\chi^2$,
a few of which are listed
in Table \ref{minimum}. With $90\%$ confidence the
neutrino mass-squared differences lie within the dots shown in
Figure \ref{islands}. Note that one of the mass-squared differences
is determined by the solar neutrino experiments and the other one
by the atmospheric neutrino observations.

If neutrinos have a hierarchy of masses (as the charged
leptons, up quarks, or down quarks), then the 
\textquotedblleft{upper island}" in Figure \ref{islands}
applies, and
$M_3 \approx 0.07$eV, $M_2 \approx 10^{-5}$eV and $M_1 < M_2$, 
with large uncertainties.

Note in
Table \ref{solar} that the ratio of the 
observed-to-predicted solar neutrino flux is significantly
lower for the Homestake experiment than for 
Sage, Gallex, Kamiokande and Super-Kamiokande which
are all compatible with $0.5$.
These latter experiments observe neutrinos
within wide energy bands, while
the chlorine detector 
in the Homestake mine observes monochromatic
electron-type neutrinos from a $^7Be$ line. 
For the Homestake experiment the spread
in $L/E$ is due to the spread in $L$, which in turn
is due to the excentricity of the orbit of the Earth.
Therefore the interference is coherent for up to 
$L \Delta M^2/(2E \cdot 2\pi) \approx 30$
oscillations from the Sun to the Earth (here $\Delta M^2$ is
either $\Delta M_{21}^2$ or $\Delta M_{32}^2$). Due to this 
coherence at \mbox{\textquotedblleft{small}"} $\Delta M^2$ it 
is possible to find acceptable
solutions with $\chi^2 < 13.4$ as shown in Figure \ref{islands}.
For larger values of $\Delta M^2$ coherence is lost and
we find solutions with $\chi^2 > 18$ which are unacceptable
if the Homestake experimental and theoretical 
errors are correct. 

\begin{table}
\begin{center}
\begin{tabular}{|l|l|}
\hline
Probability & $L/E$ [km/GeV] \\ 
\hline
$P(\bar{\nu}_e \rightarrow \bar{\nu}_e)>0.99$ & $88$ \\
$P(\nu_\mu \rightarrow \nu_\mu)>0.99$ & $0.34$ \\
$P(\nu_\mu(\bar\nu_\mu) \rightarrow \nu_e(\bar\nu_e))<0.90 \cdot 10^{-3}$ & $0.31$ \\
$P(\nu_\mu \rightarrow \nu_\tau)<0.002$ & $0.039$ \\
$P(\nu_\mu \leftrightarrow \nu_\tau) < 0.35$ & $31$ \\
$P(\nu_\mu \rightarrow \nu_\tau)<0.022$ & $0.053$ \\
$P(\nu_e \rightarrow \nu_\tau)<0.125$ & $0.031$ \\
$P(\nu_e \leftrightarrow \nu_\mu)<0.25$ & $40$ \\
$P(\bar{\nu}_e \rightarrow \bar{\nu}_e)>0.75$\cite{PaloVerde} & $232$ \\
\hline
\end{tabular}
\end{center}
\caption{Limits on the mixing probabilities from
astrophysical, accelerator and reactor experiments.\cite{rev_part_phys}}
\label{reactor}
\end{table}

\begin{table}
\begin{center}
\begin{tabular}{|l|l|l|l|l|l|l|}
\hline
$\chi_8^2$ & $M_2^2-M_1^2$ [eV$^2$] & $M_3^2-M_2^2$
[eV$^2$] & $s_{23}$ & $s_{13}$ & $s_{12}$ & $\delta$ \\
\hline
$7.0$ & $4.9 \cdot 10^{-11}$ & $5.0 \cdot 10^{-3}$ & $0.83$ & $0.08$ &
$0.50$ & $-3.14$ \\
$7.2$ & $1.6 \cdot 10^{-10}$ & $5.0 \cdot 10^{-3}$ & $0.57$ &
$0.00$ & $0.74$ & $-3.14$ \\
$7.2$ & $4.3 \cdot 10^{-10}$ & $5.0 \cdot 10^{-3}$ & $0.57$ & $0.00$ &
$0.74$ & $3.14$ \\
\hline
\end{tabular}
\end{center}
\caption{Parameters at local minima of $\chi^2$ for $8$ degrees
of freedom.}
\label{minimum}
\end{table}

\begin{figure}
\begin{center}
\vspace*{-4.5cm}
\scalebox{0.4}
{\includegraphics[0in,1in][8in,9.5in]{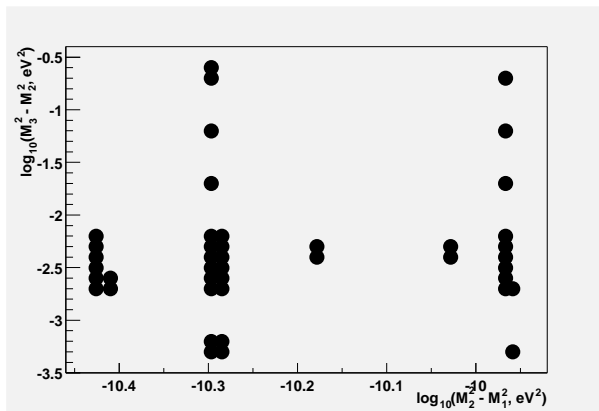}}
\vspace*{0.7cm}
\caption{Detail of the \textquotedblleft{upper island}"
of Figure \ref{islands} for the fit with $116$ degrees of
freedom (see text) at $90\%$ confidence level. The 
\textquotedblleft{lower island}" is symmetrical. 
The vertical bands correspond, from left to right, 
to $2.5$, $3.5$, $4.5$, $6.5$ and $7.5$ oscillations
from Sun to Earth of the $^7Be$ line.}
\label{116}
\end{center}
\end{figure}

An important
test of the model would be to observe seasonal variations of the
neutrino flux of the $^7Be$ line. If the lower ratio
measured by the Homestake experiment is real, 
we expect that the electron-type neutrino flux of
the $^7Be$ line is near a minimum of the 
oscillation at the average Sun-Earth distance.
In other words, there are an
odd number of half-wavelengths from Sun to Earth. 
Then we expect a modulation of the 
$^7Be$ neutrino flux with a period of half a year,
with maximums occurring at the perihelion and
aphelion of the Earth orbit.
We see no statistically significant Fourier 
component of the time dependent Homestake
data from 1970.281 to 1994.388.\cite{homestake94}
In particular the amplitude relative to the mean
of a Fourier component of period $0.5$
years is $0.09 \pm 0.10$. This observation 
implies that there are $\leq 8.5$ periods of
oscillation from Sun to Earth at $90\%$
confidence level. With a $\chi^2$ with 116
degrees of freedom, including the 8 discussed earlier
plus the 108 measurements by the Homestake
Collaboration from 1970.281 to 
1994.388\cite{homestake94} we obtain 
the allowed region shown in Figure \ref{116}.

\begin{figure}
\begin{center}
\vspace*{-4.5cm}
\scalebox{0.4}
{\includegraphics[0in,1in][8in,9.5in]{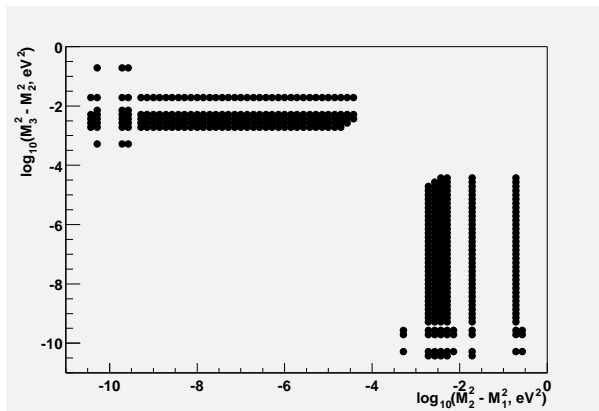}}
\vspace*{0.7cm}
\caption{Same as Figure \ref{islands} but we have
doubled the error of the Homestake experiment,
\textit{i.e.} $R=0.33 \pm 0.10$.}
\label{islands2}
\end{center}
\end{figure}

The reliability of $M_2^2 - M_1^2$ depends on the
correctness of the error assigned to the Homestake
observed-to-predicted flux ratio. 
For example, if the Homestake error listed
in Table \ref{solar} is doubled we obtain the solutions 
shown in Figure \ref{islands2}.

In view of the preceeding results let us assume
that neutrinos indeed have mass. The question
then arizes wether neutrinos are distinct from
antineutrinos (Dirac neutrinos) or wether
neutrinos are their own antiparticles
(Majorana neutrinos). This latter possibility
arizes because neutrinos have no electric
charge.

Let us consider Big-Bang nucleosynthesis
that determines the abundances of the light elements   
$D$, $^3He$, $^4He$ and $^7Li$. These abundances
are determined by the temperatures of freezout
$T_f \approx 1MeV$ when
the reaction rates $\propto T_f^5$ become comparable to the
expansion rate
$\propto T_f^2 \times (5.5 +\frac{7}{4} N_\nu)^{1/2}$.
Here $N_\nu$ is the equivalent number of massless
neutrino flavors that are ultrarelativistic at $T_f$
and are still in thermal equilibrium with photons 
and electrons at that temperature.
The calculated abundances of the light elements are in
agreement with observations
if $1.6 \le N_\nu \le 4.0$ at $95\%$ 
confidence level.\cite{rev_part_phys} 
For three generations of Majorana neutrinos, $N_\nu=3$.
For three generations of Dirac neutrinos, $N_\nu=6$
while in thermal equilibrium. However, in the
Standard Model only the left-handed 
component of neutrinos couple to $Z$, $W^+$ and $W^-$.
Right-handed neutrinos
are not in thermal equilibrium at $T_f$: their temperature
has lagged below the temperature of photons due to
the anihilation of particle-antiparticle pairs after
the decoupling of the right-handed neutrinos.
Therefore for Dirac neutrinos at $T_f$ we have 
$N_\nu \approx 3$. So we can not 
distinguish Dirac from Majorana
neutrinos using available data on nucleosynthesis.

In conclusion, the minimal extension of the Standard Model
with three massive Majorana 
or Dirac neutrinos that mix is
in good agreement with all experimental constraints.
However, confirmation of the model is needed, \textit{e.g.}
by the observation of seasonal variations of
the $^{7}Be$ spectral lines with a period of $0.5$
years, or spectral distortions and seasonal variations
of the low energy neutrinos from the solar $pp$ reaction.

\end{document}